\documentclass[letter,11pt]{article}
\pdfoutput=1
\usepackage{jheppub}
\usepackage{multirow}

\def\eps{\varepsilon}

\def\3g{{\gamma\gamma\gamma}}

\title{NNLO QCD corrections to three-photon production at the LHC}

\author[a]{Herschel A. Chawdhry,}
\author[b]{Micha\l{}  Czakon,}
\author[a]{Alexander Mitov,}
\author[a]{Rene Poncelet}

\affiliation[a]{Cavendish Laboratory, University of Cambridge, Cambridge CB3 0HE, UK}
\affiliation[b]{Institut f\"ur Theoretische Teilchenphysik und Kosmologie,
RWTH Aachen University, D-52056 Aachen, Germany}

\note{Preprint: Cavendish-HEP-19/17, TTK-19-45, P3H-19-041}

\abstract{We compute the NNLO QCD corrections to three-photon production at the LHC. This is the first NNLO QCD calculation for a $2\to 3$ process. Our calculation is exact, except for the scale-independent part of the two-loop finite remainder which is included in the leading color approximation. We estimate the size of the missing two-loop corrections and find them to be phenomenologically negligible. We compare our predictions with available 8 TeV measurement from the ATLAS collaboration. We find that the inclusion of the NNLO corrections eliminates the existing significant discrepancy with respect to NLO QCD predictions, paving the way for precision phenomenology in this process.}

\begin{document} 
\maketitle
\flushbottom

\section{Introduction}\label{sec:intro}

Over the last decade, Next-to-Next-to Leading Order (NNLO) QCD calculations for hadron collider processes have sustained tremendous progress. Owing to the development of many independent approaches \cite{Hamberg:1990np,Binoth:2000ps,Harlander:2002wh,Anastasiou:2002yz,Anastasiou:2003gr,Anastasiou:2003ds,Daleo:2006xa,Catani:2007vq,Czakon:2010td,Czakon:2011ve,Czakon:2014oma,Boughezal:2013uia,Currie:2013dwa,Boughezal:2015dva,Cacciari:2015jma,Boughezal:2016wmq,Caola:2017dug,Grazzini:2017mhc,Caola:2019nzf,Somogyi:2005xz,GehrmannDeRidder:2005cm,Abelof:2011jv,Currie:2013vh,Gaunt:2015pea,Bonciani:2015sha,Currie:2017eqf,Herzog:2018ily,Magnea:2018hab,Magnea:2018ebr,Catani:2019hip,Caola:2019pfz} almost all non-loop-induced $2\to 1$ and $2\to 2$ processes have now been computed, typically in more than one computational approach. Such massive theoretical progress has led to the creation of public codes and has started to produce valuable and solid LHC phenomenology on a massive scale. 

The computation of $2\to 3$ hadron collider processes represents a natural step beyond the current state of the art in NNLO calculations. Since many of the available computational approaches are generic, in principle they should be able to handle the problem of double real radiation in $2\to 3$ processes. The calculation of the so-called real-virtual correction to such processes should also be possible, in principle, due to the existence of numerically stable libraries for one-loop amplitudes. The only ingredient for such calculations which is not readily available are the two-loop five-point amplitudes. Thanks to the development of various new methods \cite{Papadopoulos:2015jft,Zeng:2017ipr,Chicherin:2017dob,Boels:2018nrr,Chicherin:2018ubl,Boehm:2018fpv,Chawdhry:2018awn,Gehrmann:2018yef,Abreu:2018rcw,Chicherin:2018mue,Kardos:2018uzy,Chicherin:2018old,Bendle:2019csk,Papadopoulos:2019iam}, first results for selected helicities, color structures or kinematics (typically Euclidean) have started to appear.  This includes 5-point amplitudes computed in QCD \cite{Badger:2013gxa,Gehrmann:2015bfy,Badger:2017jhb,Abreu:2017hqn,Badger:2018gip,Abreu:2018jgq,Badger:2018enw,Abreu:2018zmy,Abreu:2019odu,Badger:2019djh,Hartanto:2019uvl}, in pure Yang-Mills \cite{Badger:2015lda,Dunbar:2016aux,Badger:2016ozq}, in supersymmetric theories \cite{Bern:2006vw,Mastrolia:2012wf,Abreu:2018aqd,Chicherin:2018yne} and in gravity \cite{Dunbar:2017qxb,Chicherin:2019xeg,Abreu:2019rpt}. In this work we calculate the two-loop amplitude $q\bar q\to \3g$ which is the first time a 5-point two-loop QCD amplitude is derived explicitly, in analytic form, in the physical region. We discuss this result at length in sec.~\ref{sec:twoloop} below.

The goal of the present work is to demonstrate the feasibility of existing calculational technologies to deal with $2\to 3$ hadron collider processes. We have decided to apply this first-ever NNLO $2\to 3$ calculation to the process $pp\to \3g+X$. Our motivation for choosing this process is twofold. First, the production of colorless final states has always occupied a special place among hadron collider processes and has been the pioneering work for both $2\to 1$ and $2\to 2$ processes. In addition, the calculation of the two-loop amplitudes is more feasible due to the smallest number of scales involved.

Second, the process $pp\to \3g+X$ is of direct phenomenological interest. The cross-section for three isolated photons at the LHC 8 TeV was measured in detail by the ATLAS collaboration \cite{Aaboud:2017lxm} (see also the earlier measurement \cite{Aad:2015bua}) and was found to be {\it significantly} above the NLO QCD prediction in a wide kinematic region. Since at NLO the theory error is completely dominated by missing higher-order terms this process represents a prime case for an NNLO QCD calculation. Indeed, we find that with the help of our calculation this discrepancy can be addressed (see sec.~\ref{sec:pheno}).

The paper is organized as follows. In sec.~\ref{sec:calc} we explain our calculation with emphasis on the derivation of the two-loop amplitude. In sec.~\ref{sec:pheno} we present our predictions for the fiducial and differential cross-section. In sec.~\ref{sec:discussion} we discuss the important question of perturbative convergence in this process. Our conclusions are summarized in sec.~\ref{sec:conclusions}.

\section{The calculation}\label{sec:calc}

In this work we follow the {\tt STRIPPER} approach \cite{Czakon:2010td,Czakon:2011ve,Czakon:2014oma} previously applied at NNLO in QCD to top-pair \cite{Czakon:2014xsa,Czakon:2015owf,Czakon:2016ckf,Czakon:2016dgf,Behring:2019iiv} and inclusive jet \cite{Czakon:2019tmo} production at the LHC. The framework is implemented in a fully-differential partonic Monte Carlo program which can calculate any infrared-safe partonic observable. The technical details about our implementation can be found in ref.~\cite{Czakon:2019tmo}.

The complete calculation converges very well in terms of phase-space integration. Not counting the CPU time needed to evaluate the two-loop finite remainder (see sec.~\ref{sec:2-loop-calc} for details), it took only about 2k CPU hours to complete. The slowest contribution (about 1k CPU hours) is the real-virtual finite contribution due to the slow evaluation of the 6-point one-loop amplitude with OpenLoops 1. That contribution, however, converges fast in terms of required phase-space points.

The ingredients needed for the present calculation are tree-level amplitudes as well as the finite remainders of one-loop and two-loop amplitudes. Their calculation is described in detail in sec.~\ref{sec:oneloop} and sec.~\ref{sec:twoloop}. Here we only point out that all required one-loop amplitudes are included exactly, with full color dependence. The finite remainder of the two-loop amplitude $q \bar q\to \3g$ is included in the leading color approximation, additionally excluding diagrams with closed fermion loops. The justification for this approximation is given in sec.~\ref{sec:twoloop} below. 

The infrared subtraction operator (sometimes called ``Z"-operator) is given in ref.~\cite{Czakon:2014oma}; its leading color approximation can be found in ref.~\cite{Abreu:2018jgq}. We work in a theory with 5 massless active quark flavors and renormalize the amplitudes accordingly. No loops with massive fermions are included in our calculation. Their effect in the context of diphoton production has been discussed in ref.~\cite{Campbell:2016yrh}.

\subsection{Tree-level and one-loop amplitudes}\label{sec:oneloop}

All tree-level diagrams are computed with the help of the library {\tt avhlib} \cite{avhlib,Bury:2015dla}. For the derivation of the two-loop finite remainder the one-loop amplitude  $q \bar q\to \3g$ is needed to order $\eps^2$ (where $d=4-2\eps$ is the space-time dimension). We have computed it following the standard Feynman diagram plus Integration-by-Parts (IBP) identities \cite{Tkachov:1981wb,Chetyrkin:1981qh} approach. All required master integrals expanded to that order in $\eps$ are available in electronic form in ref.~\cite{Abreu:2018zmy}. The finite remainders for all one-loop amplitudes are obtained from the library {\tt OpenLoops} \cite{Cascioli:2011va,Buccioni:2019sur}, while the one-loop squared $q \bar q\to \3g$ contribution is taken from the library {\tt Recola} \cite{Actis:2016mpe}.

Unlike the case of diphoton production, the gluon-initiated one-loop amplitude $gg\to \3g$ vanishes and thus does not contribute to the process studied in this paper. Since the $gg$-flux is sizable, the vanishing of this contribution is of phenomenological significance and we discuss it in more detail in sec.~\ref{sec:pheno}.

\subsection{The two-loop amplitude for $q \bar q\to \3g$}\label{sec:twoloop}

An important novelty in this work is the calculation of the two-loop amplitude for the process $q \bar q\to \3g$. Although our calculation is restricted to the leading color approximation, this is the first time a two-loop five-point amplitude is put in a form that can be used in a phenomenological application. For this reason we describe it in detail in this section.

\subsubsection{Structure of the two-loop amplitude}\label{sec:structure}

We need the two-loop amplitude $|M^{(2)}(q \bar q\to \3g)\rangle$, multiplied by the Born one $|M^{(0)}(q \bar q\to \3g)\rangle$, and summed/averaged over helicities and color. Its color decomposition reads:
\begin{eqnarray}
\overline{\sum} \; 2{\rm Re}\langle M^{(2)}|M^{(0)}\rangle &=& M^{(\rm lc,\, 1)}\left(N_c^3-2N_c+1/N_c\right) +M^{(\rm lc,\, 2)}\left(N_c^3-N_c\right) \nonumber\\
&& + M^{(\rm f)}\left(N_c^2-1\right) + M^{(\rm np)}\left(N_c-1/N_c\right)\,,
\label{eq:amp}
\end{eqnarray}
where $N_c=3$ is the number of colors. 

In this work we simplify the calculation by utilizing the following approximation:
\begin{eqnarray}
\overline{\sum} \; 2{\rm Re}\langle M^{(2)}|M^{(0)}\rangle \approx N_c^3\left( M^{(\rm lc,\, 1)}+M^{(\rm lc,\, 2)}\right)\,,
\label{eq:amplc}
\end{eqnarray}
i.e. we neglect the non-planar contribution $M^{(\rm np)}$ as well as all contributions $M^{(\rm f)}$ with a fermion loop (both planar and non-planar).

The non-planar contribution $M^{(\rm np)}$ is suppressed by a factor of $1/N_c^2$ relative to the leading color one. It is thus expected to be numerically subdominant. The non-planar contribution cannot be computed at present since the required IBP solutions (topologies B$_1$ and B$_2$ in the notation of ref.~\cite{Chawdhry:2018awn}) are not yet fully known. 

The contribution $M^{(\rm f)}$ contains all diagrams with one closed fermion loop. Both planar and non-planar diagrams contribute to it. It cannot be currently derived since the required non-planar IBPs (topology B$_2$ from ref.~\cite{Chawdhry:2018awn}) are not yet known. The term $M^{(\rm f)}$ is suppressed with respect to the leading color one by a single power $1/N_c$. At the same time some diagrams
\footnote{These are the diagrams that involve no photon coupling to the closed fermion loop.}
are enhanced by the number of massless fermion flavors $n_f=5$. Therefore, although in the strict $N_c\to\infty$ limit $M^{(\rm f)}$ is suppressed relative to the leading color contribution, its numerical value may not necessarily be sub-dominant with respect to eq.~(\ref{eq:amplc}). For this reason, to be conservative, one should assume that it is comparable numerically to the leading color contribution. As we show in sec.~\ref{sec:pheno} below (see in particular fig.~\ref{fig:composition}), the numerical impact of the leading color approximation eq.~(\ref{eq:amplc}) to the differential cross-section is itself negligible, at the percent level, which {\it aposteriori} justifies the approximation $M^{(\rm f)}\approx 0$. In the future, once the corresponding contribution $M^{(\rm f)}$ becomes known (same for the term $M^{(\rm np)}$), we can easily update our cross-section predictions.

\subsubsection{Calculation of the two-loop amplitude}\label{sec:2-loop-calc}

To compute the two-loop amplitude we use a standard Feynman diagram-based approach. The diagrams are generated with the help of a private software. After multiplying with the Born amplitude and then computing the traces of spin tensors and color factors, the resulting scalar integrals are mapped to master integrals using the IBP results of ref.~\cite{Chawdhry:2018awn}. 

The last step is the inserting of the results for the required master integrals. To that end we utilize the results of ref.~\cite{Gehrmann:2018yef} where a set of integrals has been explicitly computed in terms of the so-called pentagon functions $f_{ij}$. This set of integrals can be algebraically related to the set of master integrals in ref.~\cite{Chawdhry:2018awn} with the help of the IBP solution derived in that latter reference. 

At this point the bare amplitude can be computed numerically using the routines for the numerical evaluation of the set of integrals provided with ref.~\cite{Gehrmann:2018yef}. We do not follow this approach here for two reasons. First, we would like to provide an explicit analytic result in terms of basic functions, like the set of pentagon functions. Second, the complete results involves not just the 61 master integrals but also many integrals that are obtained from them by crossings of external legs. In practice we have more than 70 set of crossings that need to be applied to the set of master integrals. While not every master integral will need to be crossed for all crossings, the complete set of integrals, accounting for all crossings, far exceeds the dimension of functions needed to describe the amplitude. For this reason such an approach would not be minimal and could lead to more severe loss of precision during the numerical evaluation.

For the above reasons, we use the explicit representation of master integrals in terms of pentagon functions $f_{ij}$ \cite{Gehrmann:2018yef} and have applied the momentum crossings directly to those functions. To minimize the set of functions, we have derived various functional identities between those functions with different arguments. As a result, we have derived an explicit expression for the squared amplitude eq.~(\ref{eq:amplc}) as a polynomial in transcendental constants and $f_{ij}$ functions with various arguments. The coefficients of this polynomial are rational functions of the kinematic invariants. They have been factorized and simplified, in some cases using the finite-field reconstruction package {\tt FiniteFlow} \cite{Peraro:2019svx}.

Besides the usual $\zeta(2)$ and $\zeta(3)$ a new set of constants, collectively called $bc4$, appear at weight 4 \cite{Gehrmann:2018yef}. Their treatments requires special attention. These constants are associated with the master integrals at weight 4 and, despite being called ``constants", in general take different numerical values in the various physical regions. We have accounted for this possibility in the process of applying momentum crossings to the master integrals. Many of these constants take the same value in the various physical regions. We have utilized the numerical values which are included in the numerical code accompanying ref.~\cite{Gehrmann:2018yef}. Similarly, the analytic continuation of the pentagon functions $f_{ij}$ across the various physical regions is performed automatically by the numerical library of ref.~\cite{Gehrmann:2018yef}. To check the correctness of our manipulations, we have compared in each physical region the numerical predictions for each master integral, constructed by us as described above, with the numerical value for the master returned directly by the library of ref.~\cite{Gehrmann:2018yef} and have found complete agreement. We have also checked many integrals against the numerical program {\tt pySecDec} \cite{Borowka:2017idc}, finding agreement in all cases.

In summary, we have expressed the complete analytic result for the bare amplitude in a basis of about 1800 transcendental terms involving $\zeta(2), \zeta(3)$ and $f_{ij}$ functions plus about 100 terms involving $bc4$ ``constants" of weight 4. 

Most of the rational coefficients are small (i.e. kB size) but some exceed 1MB. The loss of numerical precision due to cancellation between the various terms is thus of particular concern. To minimize such cancellation we have evaluated all rational coefficients with exact rational arithmetic. Specifically, we rationalize each phase-space point by preserving the accuracy of the original floating point number, and then use its rational form to compute the rational coefficients as exact rational numbers. This is implemented with the help of the {\tt CLN} library \cite{CLN}. The evaluation is much slower than the evaluation in double precision, however the overall timing is negligible compared to the evaluation of the slowest pentagon functions of weight 4. We have performed various tests for the depth of numerical cancellations and have found them to be under control in all test cases.

The numerical evaluation of the functions $f_{ij}$ is performed with the help of the {\tt C++} library provided with ref.~\cite{Gehrmann:2018yef}. The time it takes to evaluate these functions depends strongly on their weight. All functions through weight 3 are standard polylogarithms and can easily be computed with full double precision in negligible time. The functions of weight 4 are the slowest and can take up to several minutes per phase-space point. Their precision is less than full double precision due to conflicting requirements of precision and speed as well as the numerical stability of the integration routines used for their calculation. With the help of extensive experimentation we have found that computing them with at least 7 significant digits is sufficient for our purposes. To test the depth of numerical cancellations we have also computed the weight 4 functions requiring 5 significant digits. This results in a finite remainder with, typically, 2 significant digits. 

It takes about 10-50 min, depending on the phase space point, to compute the finite remainder in a single phase-space point on a single CPU (i.e. without any parallelization). The average time is about 17 min when a relative precision of $10^{-7}$ for the weight 4 functions is requested. In general, several hundred thousand points are required in order to integrate the three-photon phase space over the required bins. Such a calculation requires significant, cluster-size computer resources. While a one-off evaluation is possible it poses non-trivial problems, especially if re-evaluation of the amplitude is needed (for example for a different setup or collider energy). To minimize this computational effort we have utilized a two-fold strategy. 

First, we have produced an optimized set of phase-space points which have been generated according to the Born cross-section. Such an approach has already been used in refs.~\cite{Bevilacqua:2010qb,Bevilacqua:2015qha,Bevilacqua:2016jfk,Jones:2018hbb,Bevilacqua:2018woc} and it allows us to obtain a good quality double-virtual contribution with a reduced number of events. In practice, we have computed 30k events. 

Second, we have utilized an approach to the implementation of two-loop finite remainder where the above mentioned 30k phase space points have been used to construct a (four-dimensional) interpolating function for the real part of the finite remainder. Constructing multi-dimensional interpolating functions is a hard problem. In our case we have used the purposely developed library {\tt GPTree} \cite{GPTree} which uses advances in machine learning to optimize the interpolation tables and to produce an estimate of the interpolation error. The output of the {\tt GPTree} library is a {\tt C++} library which is portable and very easy to link to a {\tt C++} code and to use. It has the advantage that if more phase-space points are computed in the future the interpolation tables can be refitted and thus be further improved. This library will be made public in a future publication. We have also found it very useful as an additional monitor for the appearance of numerical instabilities. 

In view of the phenomenological application of the two-loop amplitude eq.~(\ref{eq:amp}), see sec.~\ref{sec:pheno}, we make explicit the scale dependence of its finite reminder ${\cal H}^{(2)}$:
\begin{eqnarray}
{\cal H}^{(2)}(\mu_R^2) ={\cal H}^{(2)}(s_{12})+\sum_{n=1}^4 c_n \ln^n\left({\mu_R^2\over s_{12}}\right)\,,
\label{eq:amp-scale}
\end{eqnarray}
where $s_{12}$ is the partonic c.m. energy squared. Since the coefficients $c_n$ can be determined exactly from the tree-level and one-loop amplitudes, throughout this work the scale dependence of the two-loop finite reminder is included with full color dependence. Therefore, the approximation eq.~(\ref{eq:amplc}) is applied only to the first term in the RHS of eq.~(\ref{eq:amp-scale}). As can be seen in fig.~\ref{fig:composition} below, the numerical impact on the NNLO cross-section of the scale-independent part of the two-loop finite remainder ${\cal H}^{(2)}(s_{12})$ is rather small, at the percent level. The explicit expressions for the amplitude, as well as further details about its evaluation, will be given in a subsequent publication.

\section{Phenomenology}\label{sec:pheno}

\subsection{LHC setup}\label{sec:setup}

Our calculational setup follows the 8 TeV ATLAS measurement \cite{Aaboud:2017lxm}. The definition of histograms and experimental data is taken from the corresponding HEPData entry \cite{hepdata}. Our event selection is based on the following phase-space cuts:
\begin{itemize}
\item $E_T$ cut for the three photons: $E_{T,\gamma_1} > 27$ GeV, $E_{T,\gamma_2} > 22$ GeV and $E_{T,\gamma_3} > 15$ GeV, where $\gamma_1$ represents the hardest photon while $\gamma_3$ the softest one.
\item All photons are required to have $|\eta_\gamma| < 2.37$, excluding the range $1.37<|\eta_\gamma|<1.56$.
\item Photon separation: the angular distance $\Delta R$ between any two photons is required to be $\Delta R_{ij}>0.45$, where $\Delta R_{ij}=\sqrt{\left(\eta_i-\eta_j\right)^2+\left(\phi_i-\phi_j\right)^2}$.
\item A minimum three-photon invariant mass is required: $m_\3g>50$ GeV.
\item Following Frixione \cite{Frixione:1998jh}, we impose smooth photon isolation. Specifically, for any angular distance $\Delta R$ away from each photon, subject to $\Delta R \le \Delta R_0$, we require

\begin{equation}
E_T^{\rm iso}(\Delta R) < E_T^{\rm max} {1-\cos(\Delta R)\over 1-\cos (\Delta R_0)}\,,
\end{equation}

where $\Delta R_0 = 0.4$ and $E_T^{\rm max} =10~{\rm GeV}$. The energy $E_T^{\rm iso}(\Delta R)$ is defined as

\begin{equation}
E_T^{\rm iso}(\Delta R) = \sum_i E_{T,i} \Theta (\Delta R - \Delta R_{i,\gamma})\,.
\end{equation}
The sum in the above equation is over all final-state partons $i$, and $E_{T,i}$ and $\Delta R_{i,\gamma}$ are parton $i$'s transverse energy and angular distance with respect to the photon.
\end{itemize}

Our calculation uses the {\tt NNPDF31\_nnlo\_as\_0118} pdf set \cite{Ball:2017nwa}. We have not computed the pdf error; it was estimated in ref.~\cite{Aaboud:2017lxm} and found to be below the (NLO) scale variation. 

In this work we have utilized two different forms for the dynamic factorization and renormalization scales:
\begin{eqnarray}
H_T &\equiv& \sum_{i=1}^3 E_{T,\gamma_i}\label{eq:scaleHT}\,,\\
M_T &\equiv& \sqrt{p_{\3g,T}^2+m_\3g^2} \quad \text{with} \quad p_\3g = \sum_{i=1}^3 p_{\gamma_i} \quad \text{and} \quad  m_\3g^2=p_\3g ^2\label{eq:scaleMT}\;.
\end{eqnarray}

Our default central scale choice is $\mu_0 = H_T/4$, which follows from the findings of ref.~\cite{Czakon:2016dgf}. In fact, in the following we have studied the choices $\mu_0 = H_T/n$, with $n=1,2,4$ as well as the alternative choices $\mu_0 = M_T/n$, with $n=1,2,4$, that are based on the transverse mass of the three-photon system. The $M_T$-based scale was used in the latest diphoton production study \cite{Catani:2018krb}. We find that the differences between calculations with central scales $M_T/n$ and $H_T/n$, with $n=1,2,4$, are relatively small. Scale variation of the factorization and renormalization scales is derived with a standard seven-point variation around the central scale $\mu_0$.

\subsection{Fiducial cross-section}\label{sec:fiducial}

\begin{figure}[t]
\includegraphics[width = 1\textwidth,trim=10mm 10mm 0 0]{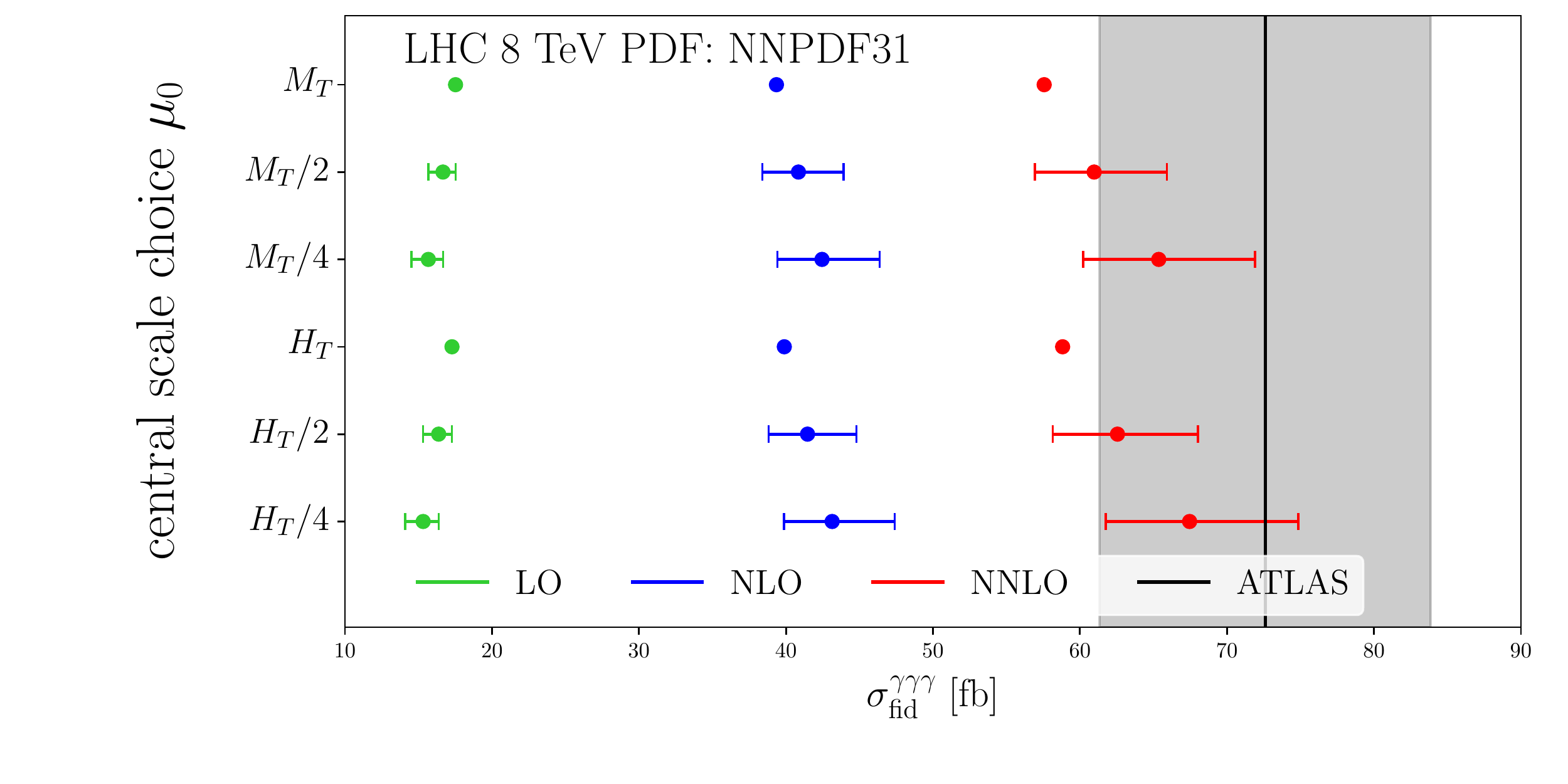}
\caption{Predictions for the fiducial cross-section in LO (green), NLO (blue) and NNLO (red) QCD versus ATLAS data (black). Shown are predictions for six scale choices. The error bars on the theory predictions reflect scale variation only. For two of the scales only the central predictions are shown.}
\label{fig:fiducial}
\end{figure}

In fig.~\ref{fig:fiducial} we compare ATLAS data \cite{Aaboud:2017lxm} with the predictions for the fiducial cross-section as defined in sec.~\ref{sec:setup}. We compare predictions based on 6 different renormalisation and factorization scales, in LO, NLO and NNLO QCD. In all cases we observe large shifts from LO to NLO and from NLO to NNLO which are much larger than the scale variations at, respectively, LO and NLO. Specifically, for our default scale $\mu_0=H_T/4$ we have an NLO/LO correction of about 2.8 while the NNLO/NLO correction is about 1.6. We discuss this important feature in sec.~\ref{sec:discussion} below.

Predictions based on the two different scale functional forms eqns.~(\ref{eq:scaleHT},\ref{eq:scaleMT}) are rather similar relative to the sizes of scale variations and experimental uncertainties. Therefore, in the following, we will mainly focus our discussion on the $H_T$-based scales. 

The scales $\mu_0=H_T/4$ and $\mu_0=H_T/2$ both agree with data, especially the $H_T/4$ one. The scale $\mu_0=H_T$ is only just outside the measurement's uncertainty band. For simplicity in this work we did not compute the full scale variation around the scale $H_T$ (same for $M_T$) which is why only the central value is shown for these two scales. We do not expect the scale variation around the two scales will be much different that the pattern already emerging from fig.~\ref{fig:fiducial}.

In general we observe that the scale variation increases when going from LO to NNLO and that all scales are consistent at a given order within their scale uncertainties. For a proper interpretation of the reliability of the theoretical predictions it is therefore imperative to understand the issue of perturbative convergence. We devote sec.~\ref{sec:discussion} to this issue but here we only say in advance that we believe the NNLO predictions are probably the first order for which the theory prediction, with its associated scale variation, is reliable. 

To summarize, based on the above discussion we conclude that our default scale choice is in perfect agreement with the experimental measurement
\begin{eqnarray}
\sigma_{\rm fid}({\rm ATLAS}) &=& 72.6 \pm 6.5 ({\rm stat.}) \pm 9.2 ({\rm syst.})\, {\rm fb}\,,\nonumber\\
&&\nonumber\\
\sigma_{\rm fid}({\rm NNLO\, QCD};~H_T/4) &=& 67.5 ^{+7.4\; (11\%)} _{-5.7\; (8\%)}\, ({\rm scales})~ {\rm fb}\,.
\end{eqnarray}
Clearly, the inclusion of the NNLO QCD correction plays a crucial role in this agreement. 

The MC error on the fiducial NNLO prediction is below 1\%. The fiducial predictions based on the various scale choices are available in electronic form with the arXiv submission of this article.

\subsection{Differential distributions}\label{sec:differential}

\begin{figure}[t]
\includegraphics[width = 0.32\textwidth,trim=0 1mm 0 0]{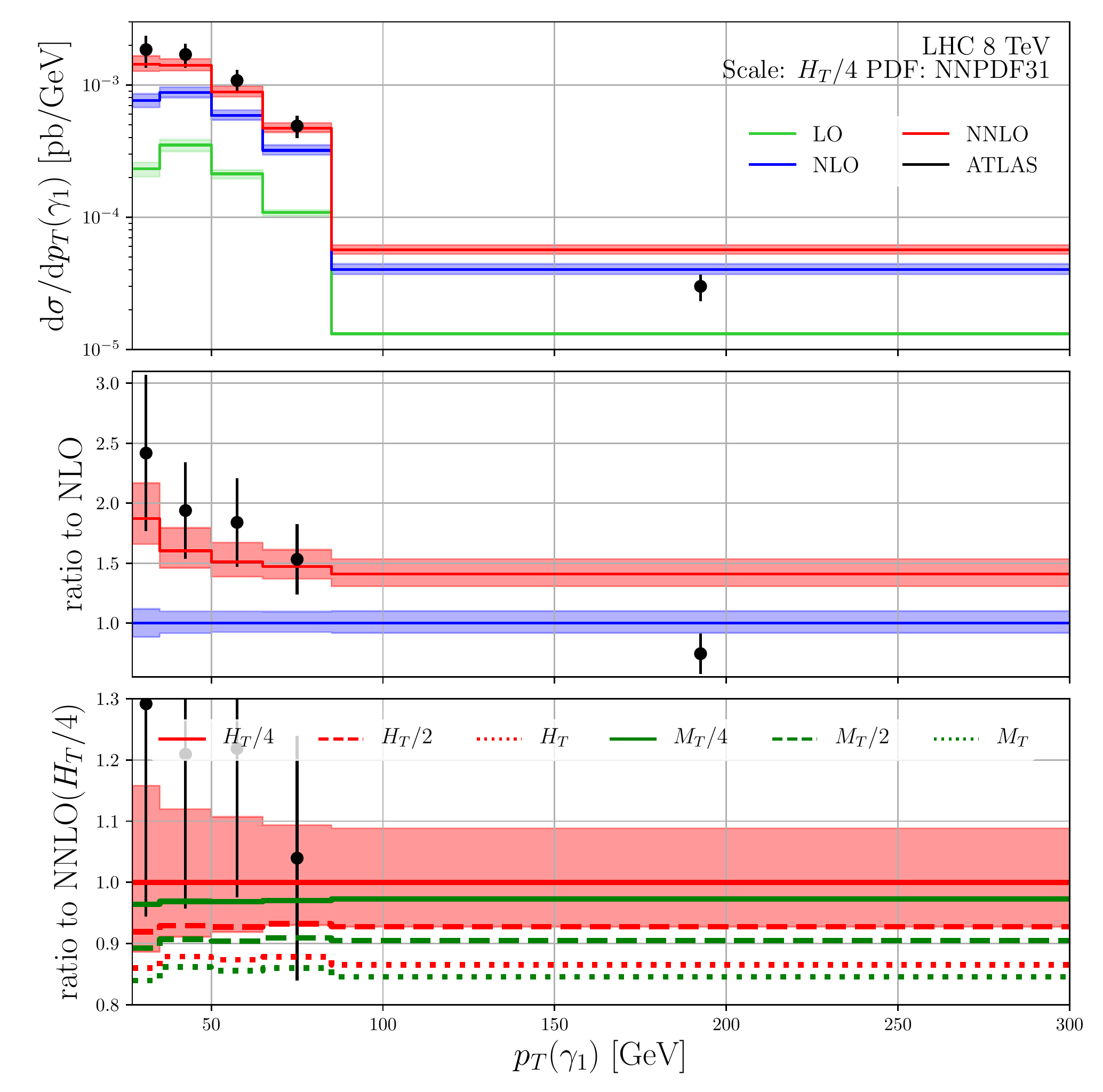}
\includegraphics[width = 0.32\textwidth,trim=0 1mm 0 0]{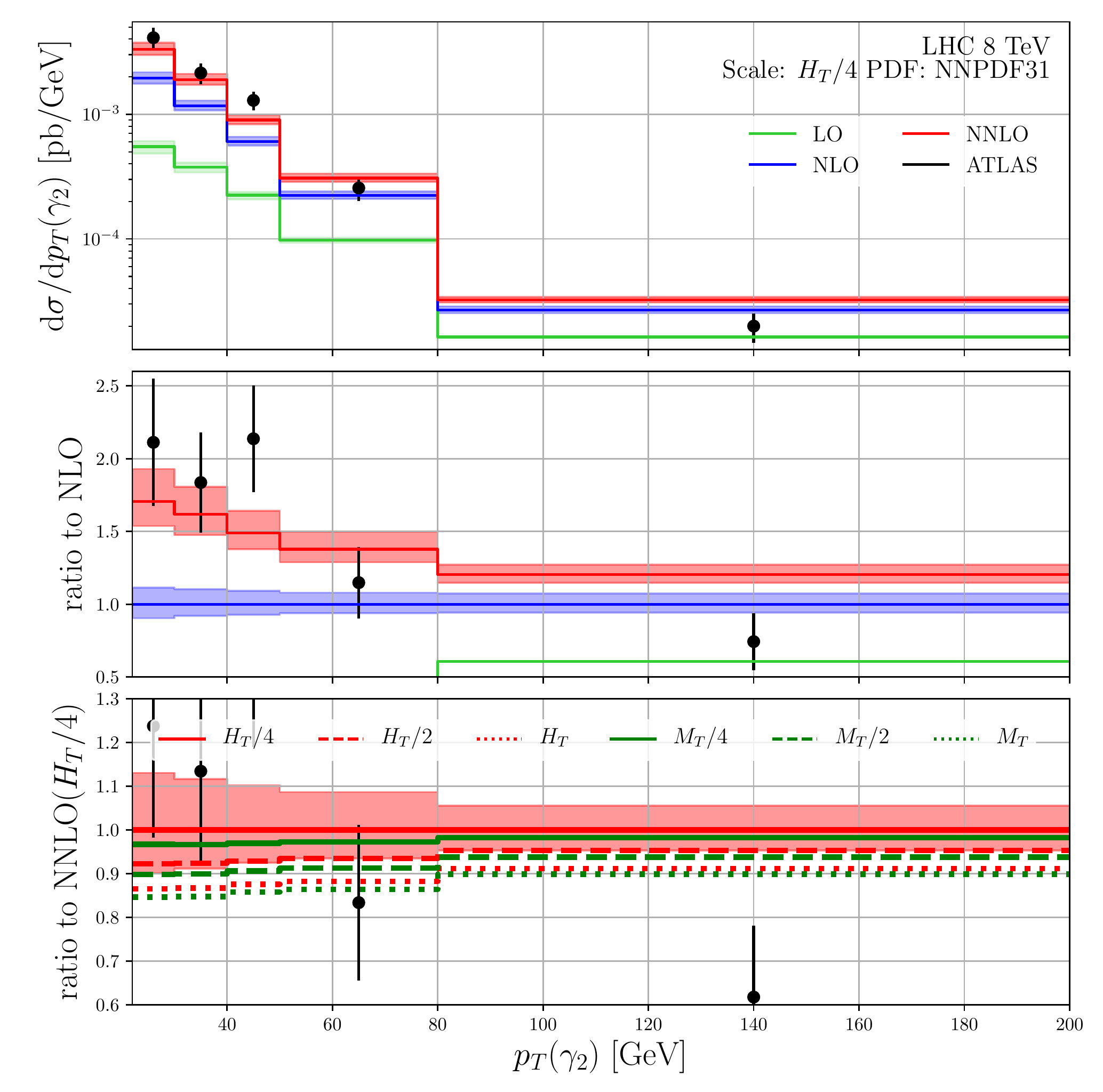}
\includegraphics[width = 0.32\textwidth,trim=0 1mm 0 0]{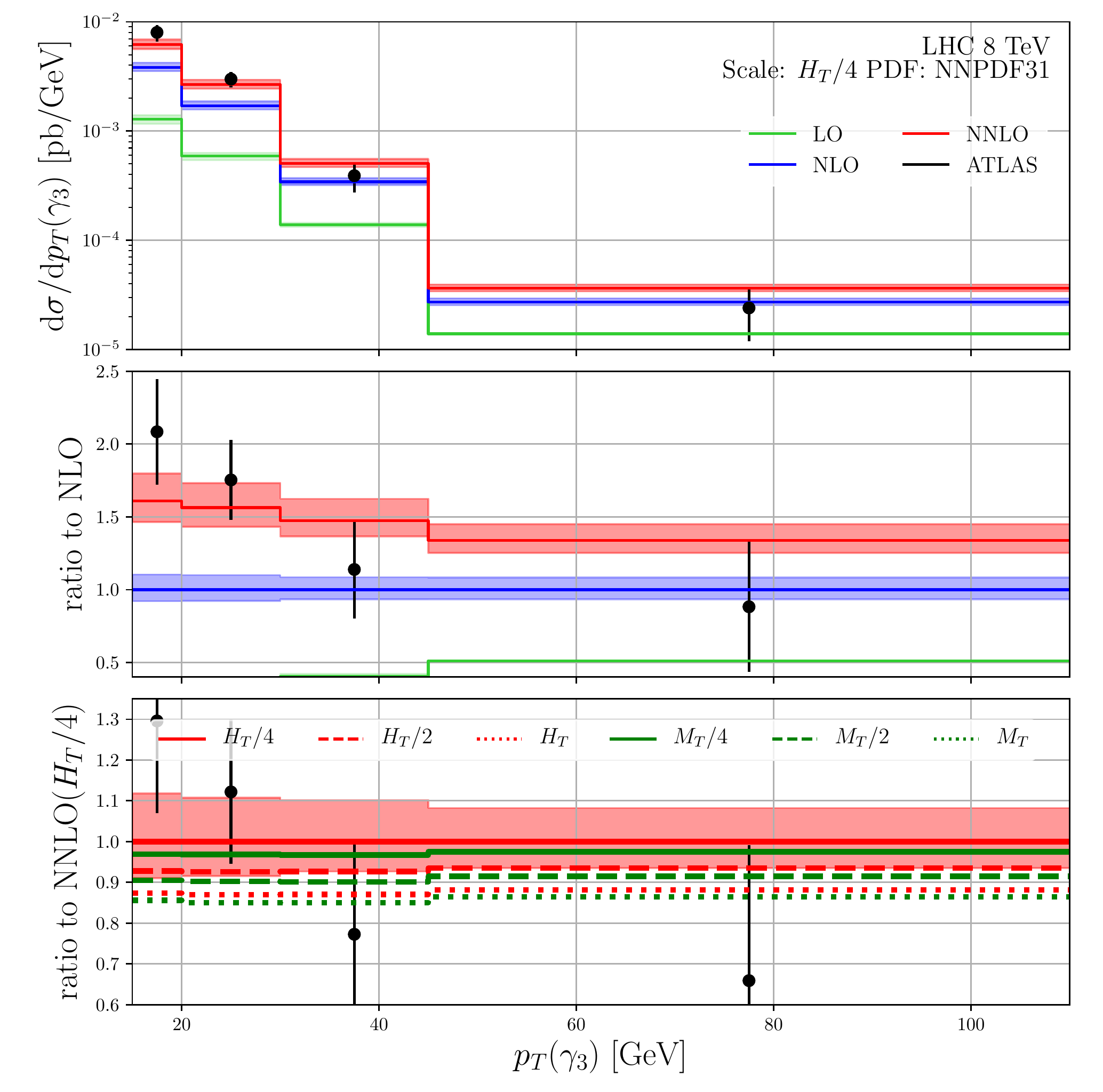}
\caption{$p_T$ distribution of the hardest photon $\gamma_1$ (left), $\gamma_2$ (center) and the softest one $\gamma_3$ (right). Top plot shows the absolute distribution at NNLO (red), NLO (blue) and LO (green) versus ATLAS data (black). Middle plot shows same distributions but normalized to the NLO. Bottom plot shows central NNLO predictions for 6 different scale choices (only the central scale is shown) with respect to the default choice $\mu_0=H_T/4$. The bands represent the 7-point scale variations about the corresponding central scales.}
\label{fig:PT}
\end{figure}
\begin{figure}[t]
\includegraphics[width = 0.32\textwidth,trim=0 1mm 0 0]{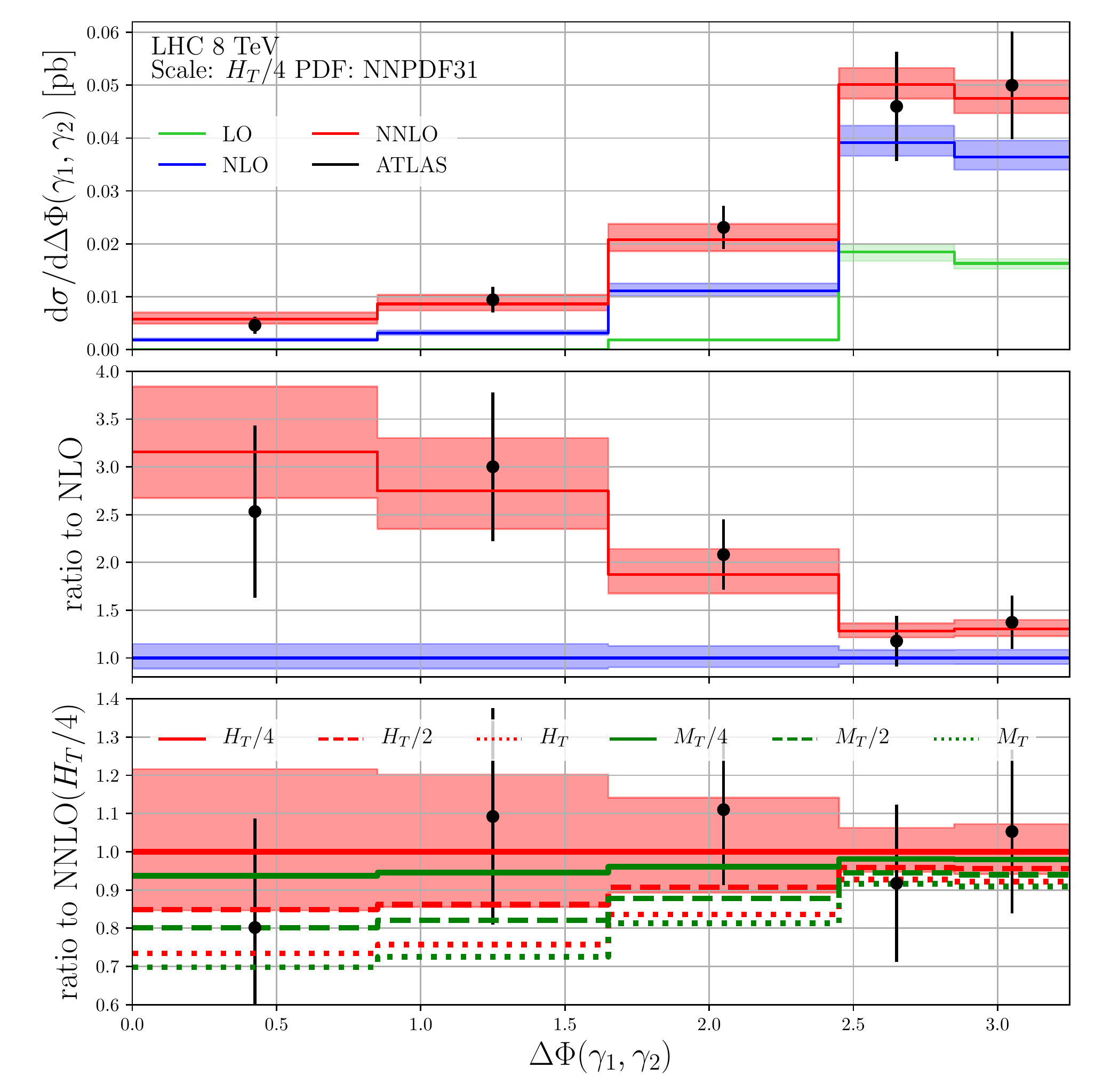}
\includegraphics[width = 0.32\textwidth,trim=0 1mm 0 0]{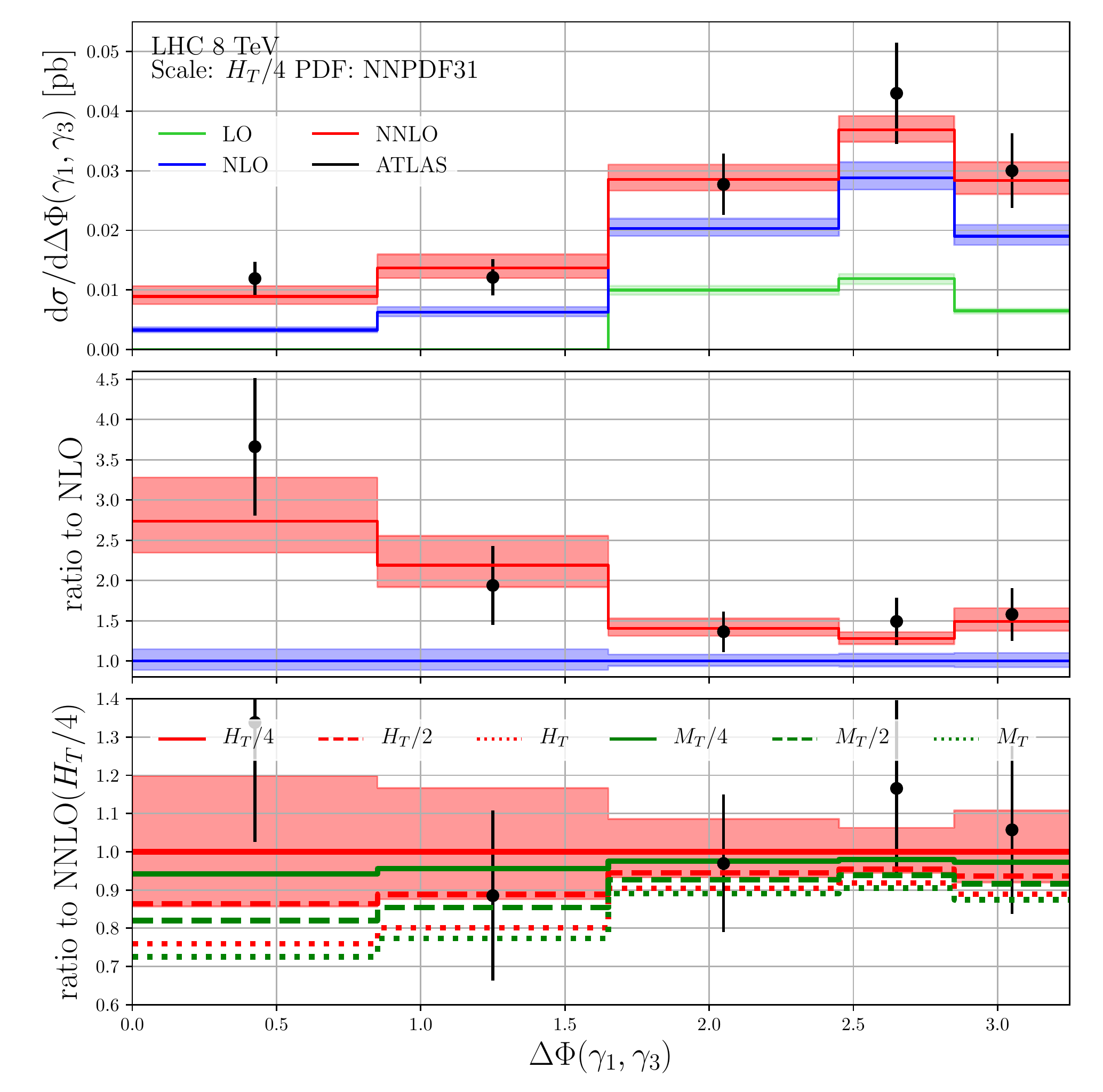}
\includegraphics[width = 0.32\textwidth,trim=0 1mm 0 0]{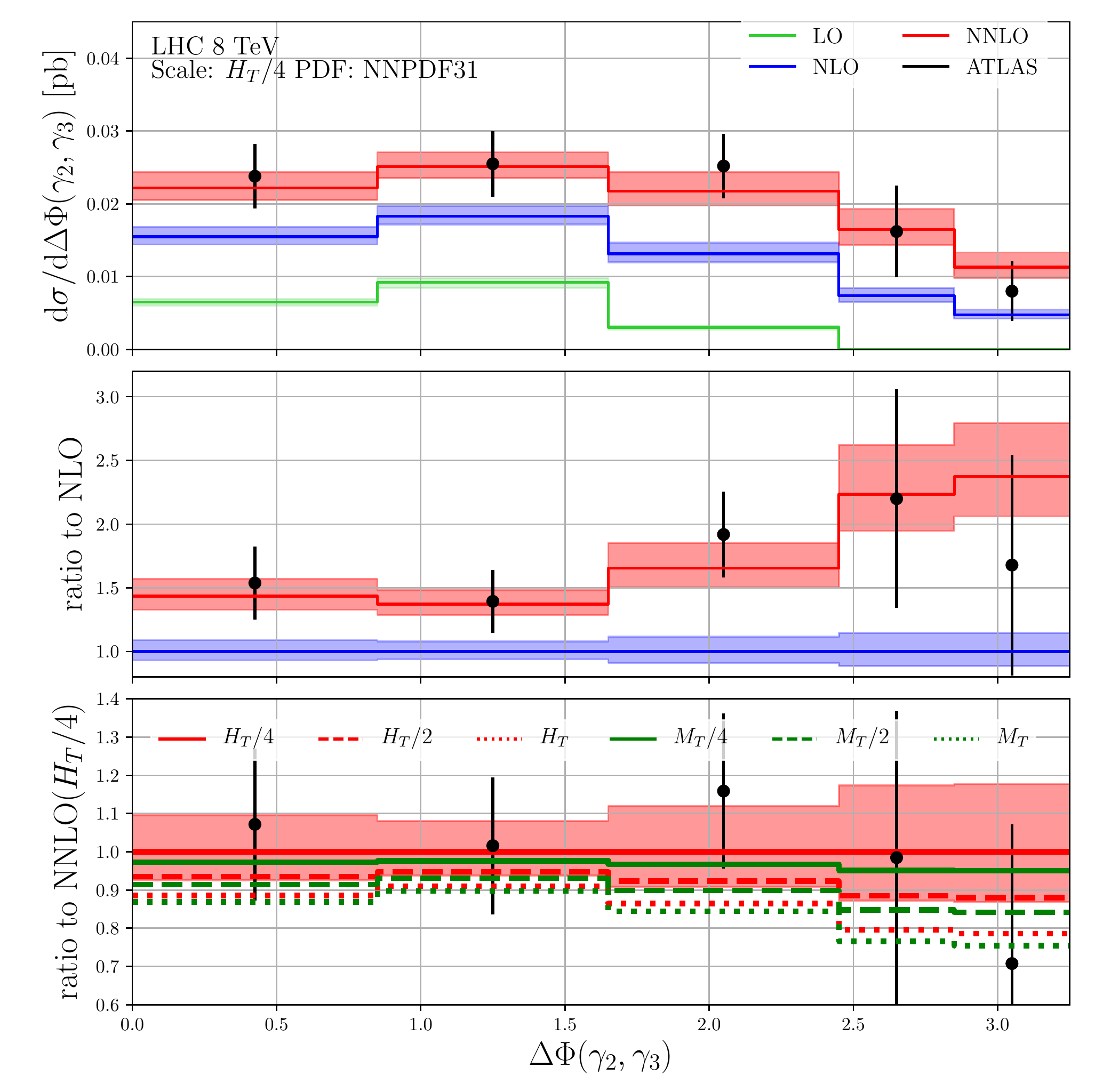}
\caption{As in fig.~\ref{fig:PT} but for the $\Delta\Phi(\gamma_i,\gamma_j)$ distributions.}
\label{fig:Dphi}
\end{figure}
\begin{figure}[t]
\includegraphics[width = 0.32\textwidth,trim=0 1mm 0 0]{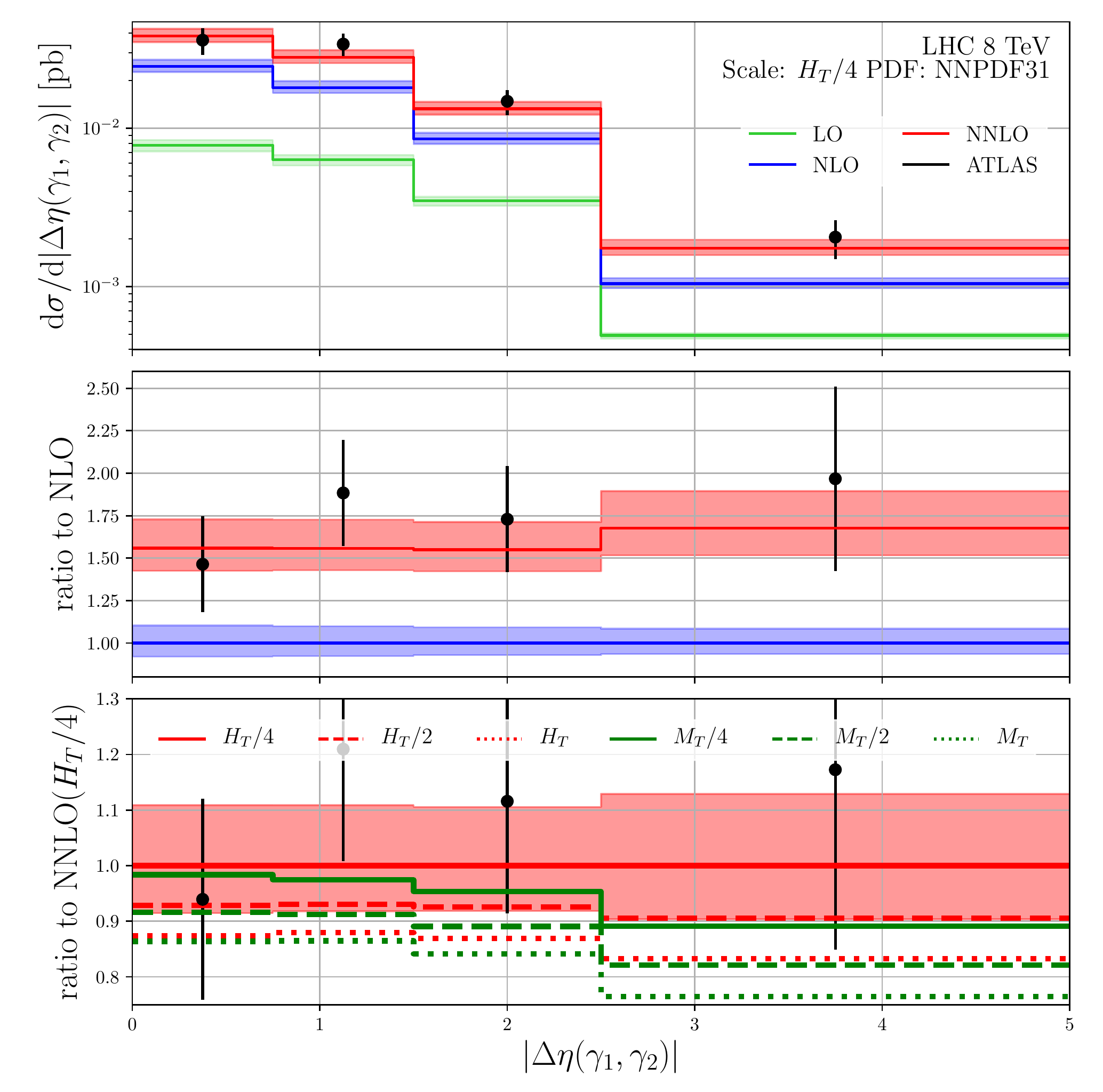}
\includegraphics[width = 0.32\textwidth,trim=0 1mm 0 0]{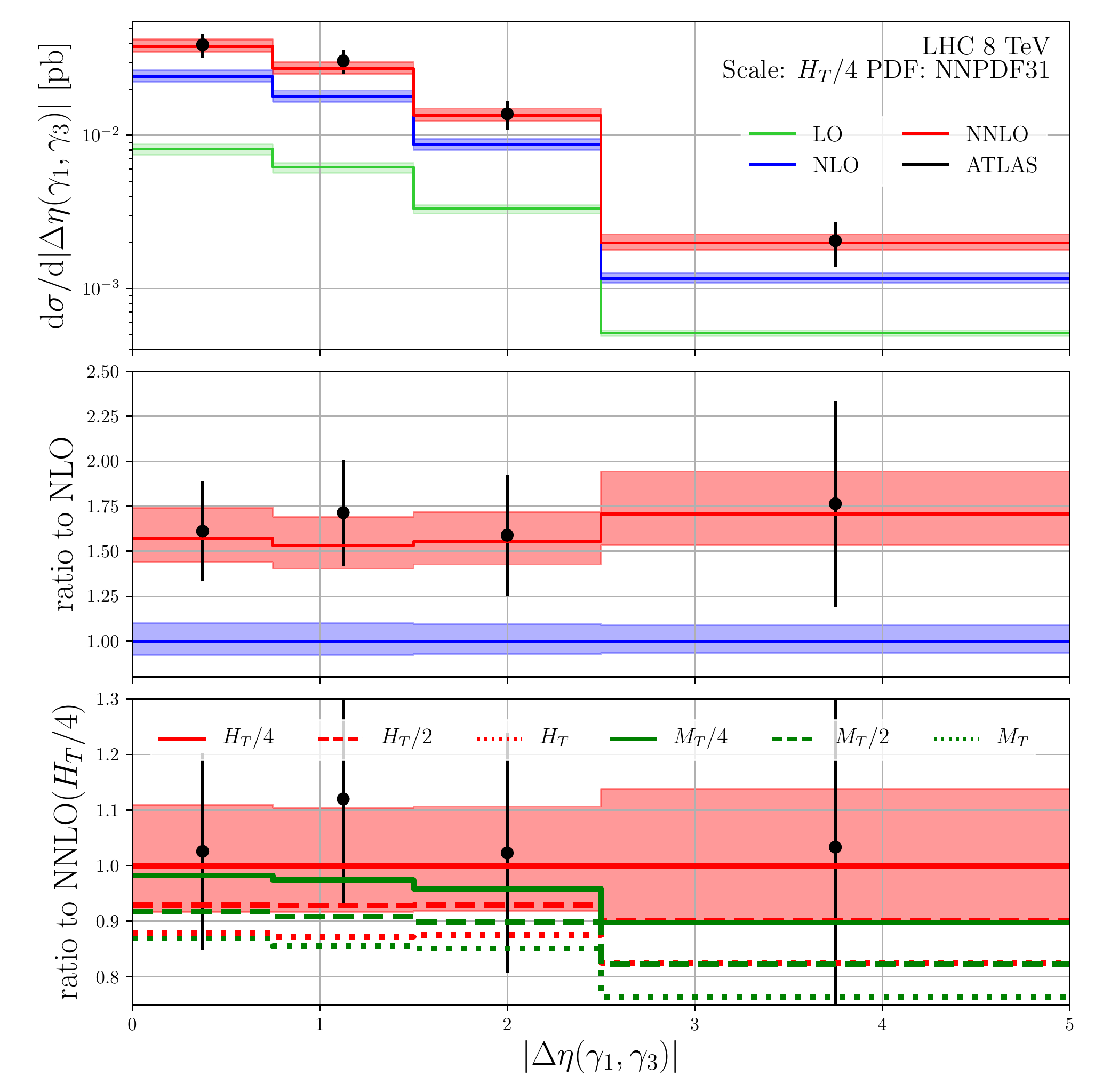}
\includegraphics[width = 0.32\textwidth,trim=0 1mm 0 0]{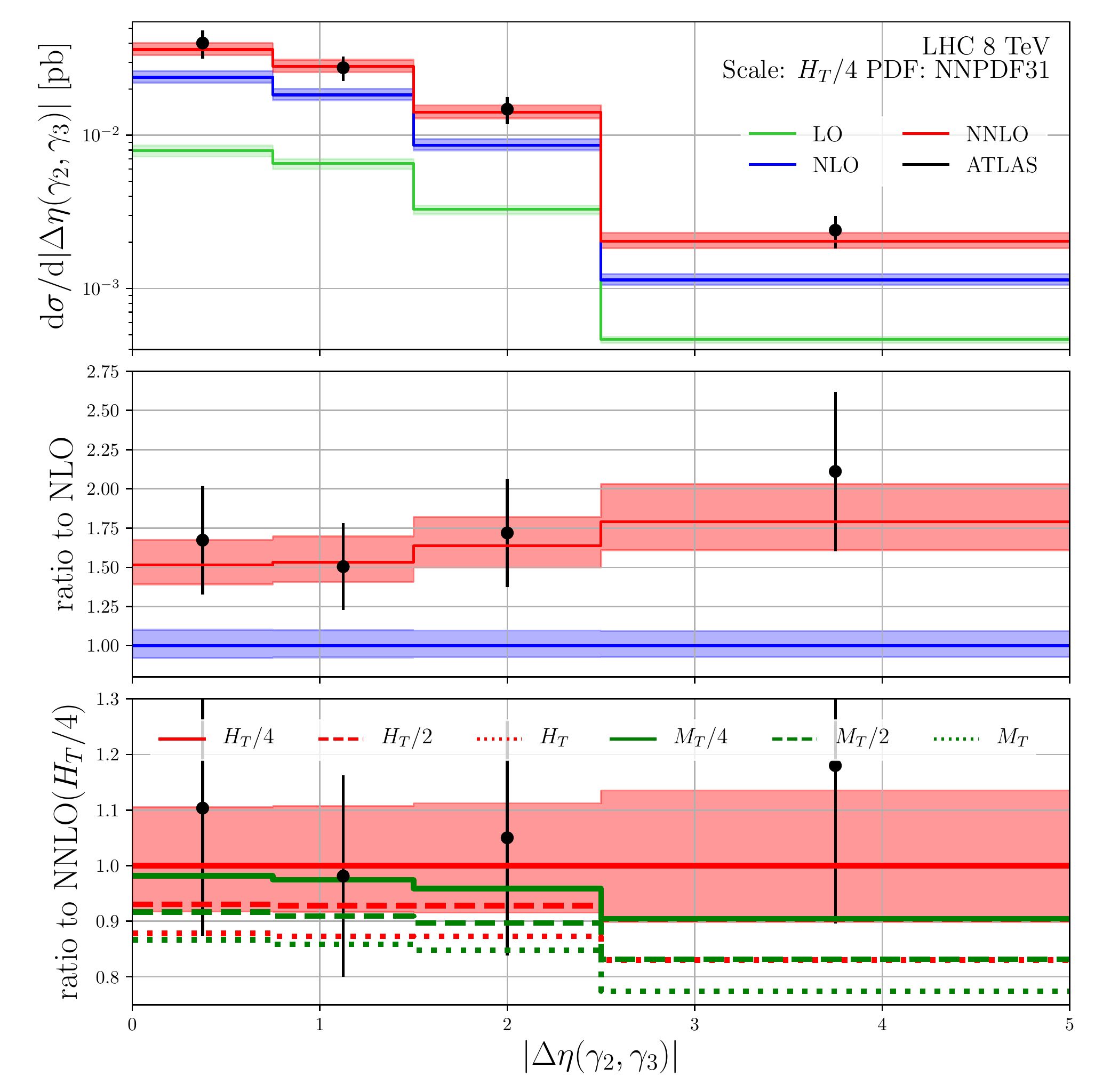}
\caption{As in fig.~\ref{fig:PT} but for the $|\Delta\eta(\gamma_i,\gamma_j)|$ distributions.}
\label{fig:Deta}
\end{figure}

A very large number of differential distributions have been measured by the ATLAS collaboration in ref.~\cite{Aaboud:2017lxm}. In this work we have computed the theory predictions in NNLO QCD for all of them. 

\begin{figure}[t]
\includegraphics[width = 0.50\textwidth,trim=0 0mm 0 0]{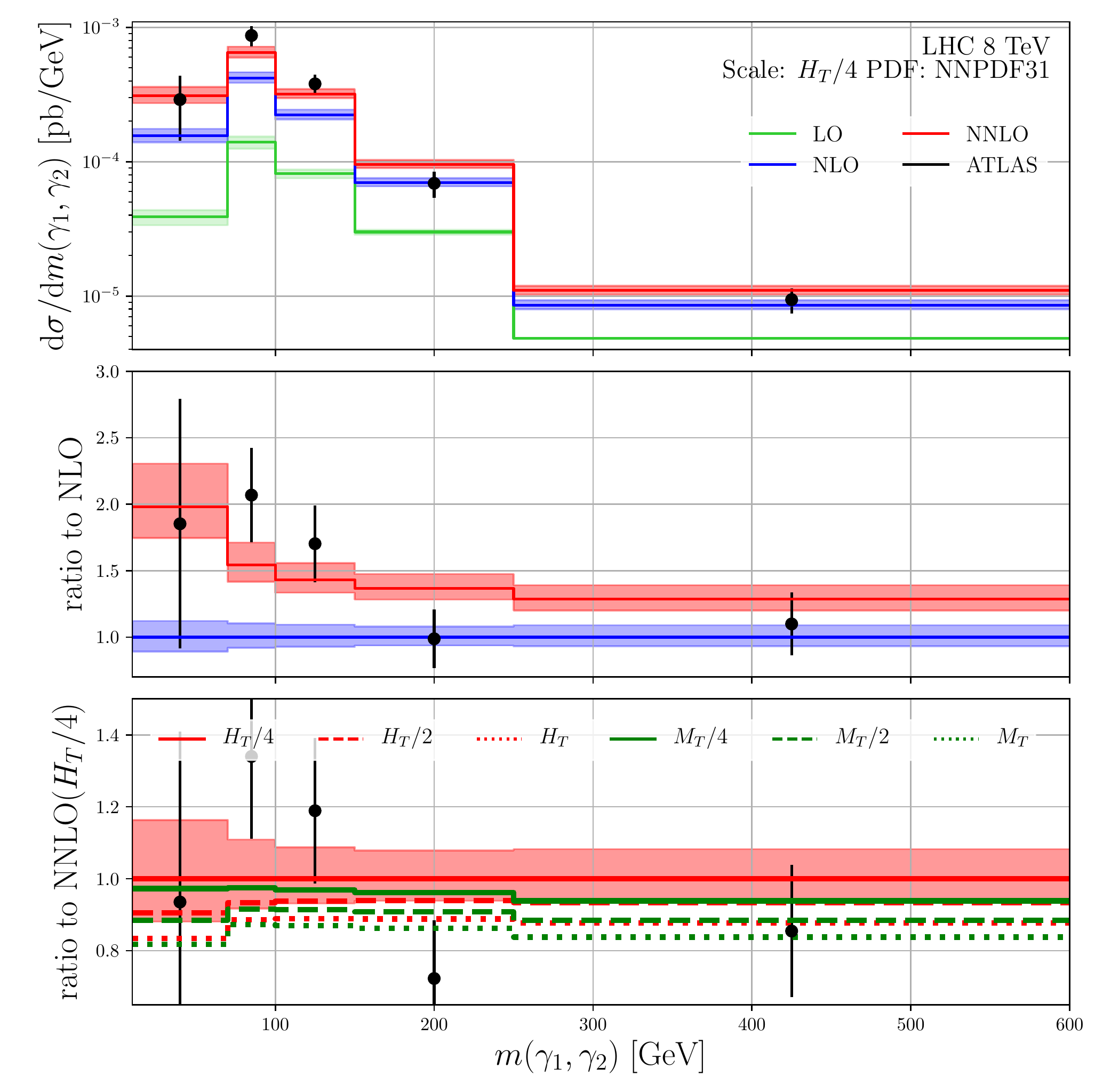}
\includegraphics[width = 0.50\textwidth,trim=0 0mm 0 0]{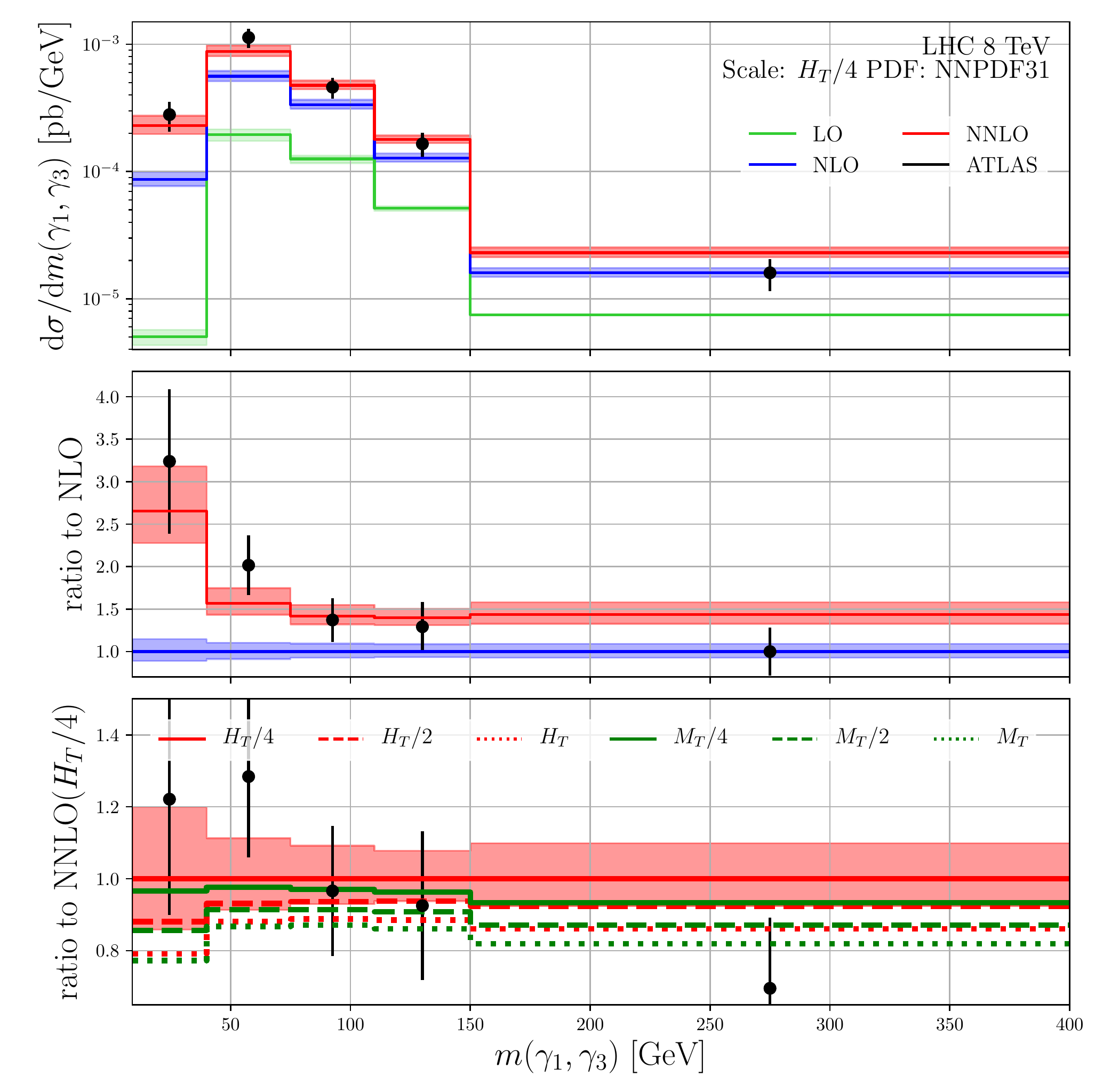}
\includegraphics[width = 0.50\textwidth,trim=0 10mm 0 0]{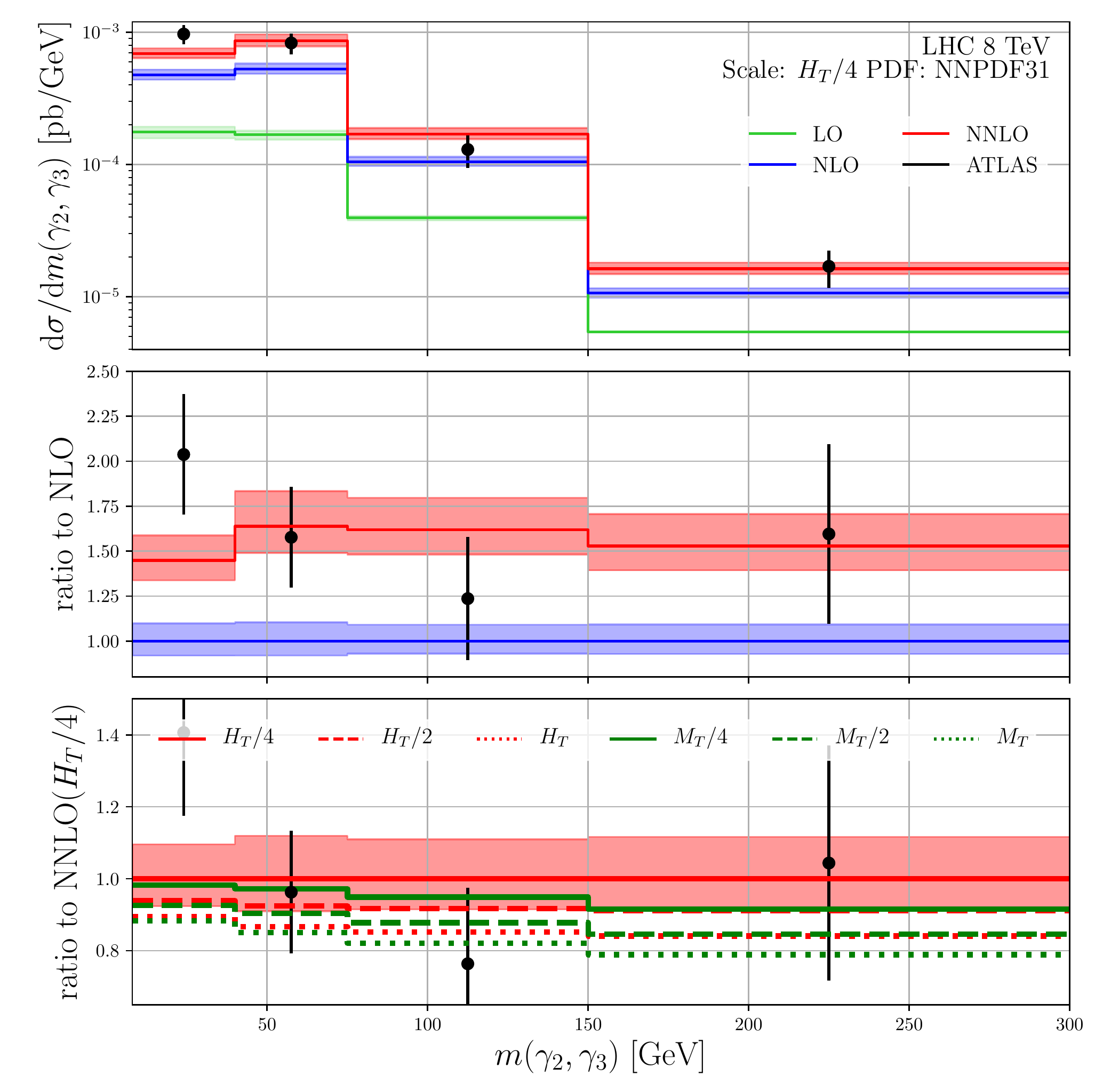}
\includegraphics[width = 0.50\textwidth,trim=0 10mm 0 0]{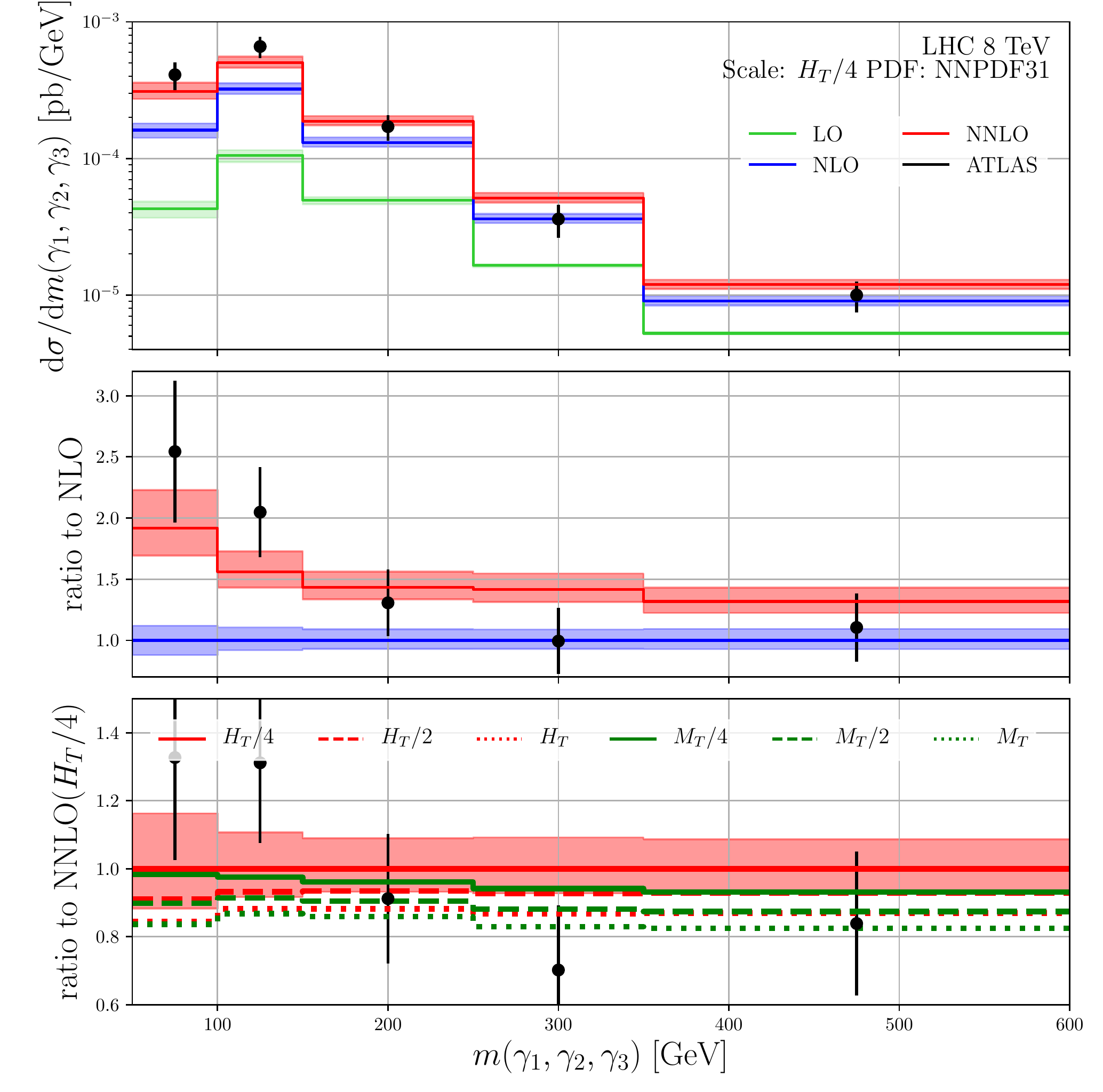}
\caption{As in fig.~\ref{fig:PT} but for the $m(\gamma_i,\gamma_j)$ and $m(\gamma_1,\gamma_2,\gamma_3)$ distributions.}
\label{fig:Mass}
\end{figure}

We start by showing in fig.~\ref{fig:PT} the predictions for the $p_T$ distributions of the three individual photons: the hardest one $\gamma_1$ (left), $\gamma_2$ (center) and the softest one $\gamma_3$ (right). We show the absolutely normalized distributions at LO (green), NLO (blue) and NNLO (red) in QCD. The top and middle panels show the central scale predictions and their corresponding 7-point scale variations for our default scale choice $\mu_0=H_T/4$:  the top panel shows the absolutely normalized distributions while the middle one shows the same results but normalized to the NLO central predictions. Shown in black is the ATLAS data. The bottom panels show the central scale predictions for the other 5 scale choices normalized to the central scale value for the default scale $H_T/4$. For a more quantitative comparison we also show the scale variation band of the default scale as well as the ATLAS data.

The plots for all other differential distributions shown next follow the same pattern: in fig.~\ref{fig:Dphi} we show the three $\Delta\Phi$ angles between the three pairs of photons, in fig.~\ref{fig:Deta} we show the three rapidity differences $|\Delta\eta|$ between the three pairs of photons and, finally, in fig.~\ref{fig:Mass} we show the invariant mass distributions between the three pairs of photons as well as the invariant mass of the three-photon system.

Overall, a very consistent picture arises from all differential distributions, both in relation to the properties of the theory predictions as well as in relation to their agreement with data.

The most notable feature evident in all differential distributions are the large jumps from LO to NLO and from NLO to NNLO. The difference between orders is much larger than the corresponding scale variations at LO and NLO which, in principle, raises the question of the validity of perturbative convergence in this process. This behavior closely resembles the one already discussed for the fiducial cross-section. At this point we will only mention that we believe the NNLO QCD predictions is likely already a reliable prediction which can be confidently compared to data. We leave the detailed discussion of this point to sec.~\ref{sec:discussion}. 

The second notable feature is the overall good agreement between NNLO QCD predictions based on a scale $H_T/4$ with data. While in most distributions there are bins that do not agree with the NNLO prediction, the overall shape as well as normalization of all distributions is clearly correctly described at NNLO. In fact in some of the bins where deviations is observed could be due to larger statistical fluctuations in data. An improved future measurement will clearly be very useful to clarify this. Interestingly, it is the distributions $\Delta\Phi$ and $\Delta\eta$ that are described best and in fact in those two we observe perfect agreement between NNLO and data for all pairs of photons. 

The relative MC error on the differential NNLO predictions shown here is below 3 percent. The theoretical predictions for all distributions, based on our default scale $H_T/4$, are available in electronic form with the arXiv submission of this article.

In summary, we would like to stress that in this calculation we have only accounted for the QCD corrections through NNLO. Other theoretical contributions should at this point also be revisited. These include  electroweak corrections and (refining the study of) effects due to photon isolation. Effects due to pdfs appear to be subdominant to the scale variation at NNLO but this should also be cross-checked in a more complete study. The issue of the ``best" scale choice can always be debated and at this level of precision seems to be a dominant source of theoretical uncertainty. For a detailed phenomenological study the MC error of the predictions shown here can be improved further. Finally, for completeness, one would like to have the complete NNLO prediction by including the contributions to the two-loop finite remainder neglected in this work, although we expect them to be phenomenologically insignificant.

\subsection{Discussion of perturbative convergence}\label{sec:discussion}

As in diphoton production \cite{Catani:2011qz,Campbell:2016yrh,Catani:2018krb}, the inclusive production of three photons exhibits behavior that at first glance is inconsistent with perturbative convergence. Indeed, as emphasized above, we observe large jumps from LO to NLO and from NLO to NNLO. These jumps are much larger than the corresponding scale variation bands at LO and NLO. This behavior is evident in all differential distributions as well as in the fiducial cross-section. Specifically, we recall that for our default scale choice the fiducial cross sections at NLO exceed the LO one by a factor of 2.8 while the NNLO/NLO K-factor is about 1.6. This behavior is very similar to the one encountered in diphoton production.

Various arguments have been given in the past for the appearance of such large K-factors in diphoton production. Two of those arguments are the presence of asymmetric cuts imposed on the two photons as well as the sizable loop-induced $gg\to\gamma\gamma$ contribution. While these arguments have their merit, it is easy to see that they are not the drivers behind the behavior we are trying to understand in diphoton (as well as three-photon) production. For example, the asymmetric cuts should not play an appreciable role for three-photon production because the Born state is naturally asymmetric. Similarly, while the loop-induced reaction is very large relative to the LO diphoton cross-section, its relative contribution at NNLO is not that sizable, only of the order of 10\% \cite{Catani:2018krb}. While such a contribution is important it is not large enough to be the driver behind the large K-factors observed in both processes. In fact, this issue can be cleanly understood in three-photon production process where the corresponding loop-induced amplitude $gg\to \3g$ vanishes.

\begin{figure}[t]
\includegraphics[width = 1\textwidth,trim=10mm 10mm 0 0]{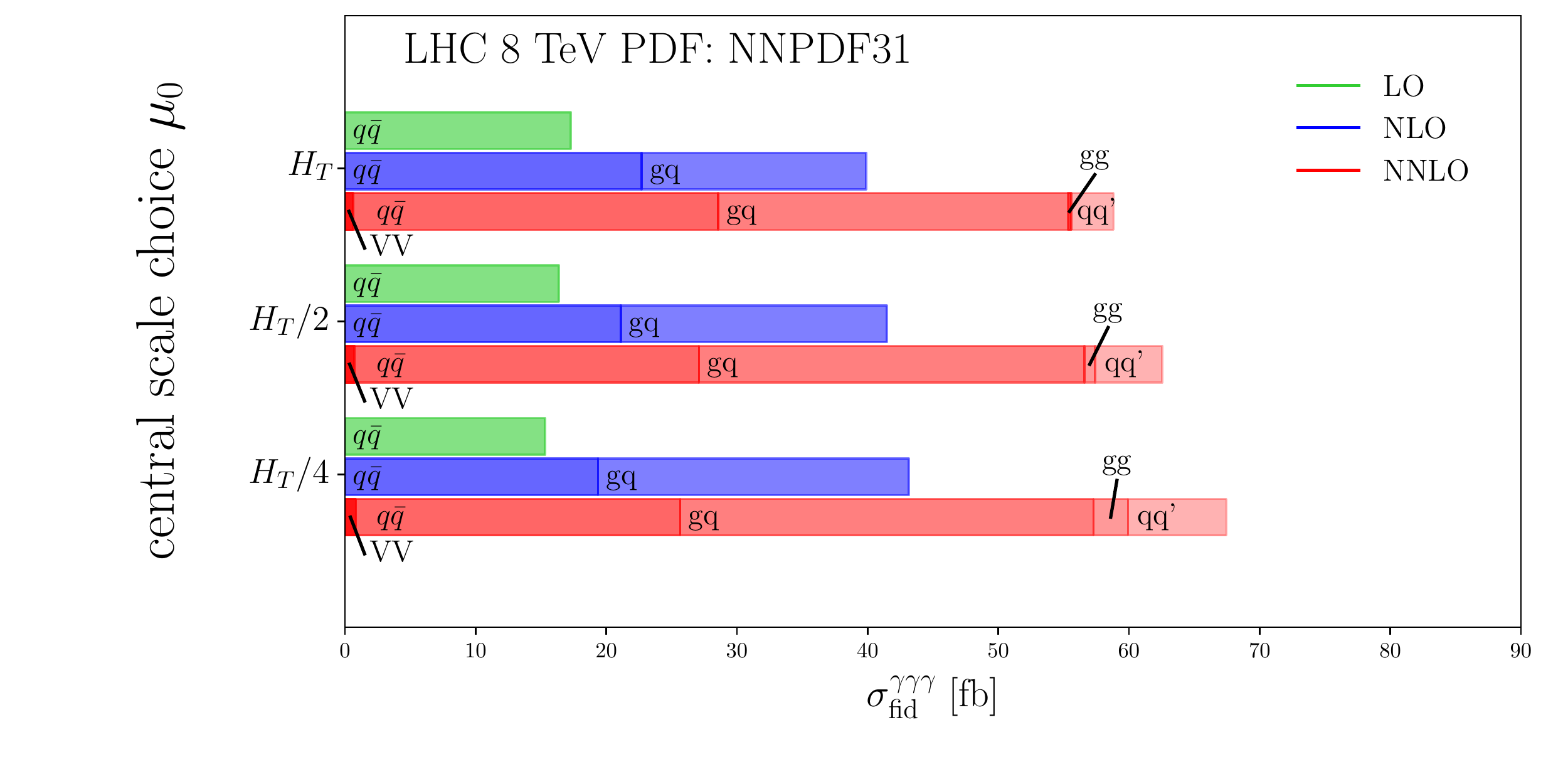}
\caption{Anatomy of higher-order corrections to the three-photon fiducial cross-section at LO (green), NLO (blue) and NNLO (red) by partonic channels for three different central scale choices. Also shown is the contribution from the scale-independent part of the two-loop finite remainder (VV) computed in our approximation defined in sec.~\ref{sec:structure}.}
\label{fig:composition}
\end{figure}

The above analysis of the $gg$-driven correction brings a very important point, namely, the role the initial-state flux plays in the apparent perturbative convergence of these two processes. To quantify this in fig.~\ref{fig:composition} we show the composition of the fiducial cross-section at LO, NLO and NNLO organized by initial-state partonic reactions. We show the results for three different $H_T$-based central scales; the results for the corresponding $M_T$-based scales are very similar.

What we observe in fig.~\ref{fig:composition} is very illuminating. First, we note that the $gg$ flux does contribute (due to double real emissions and collinear subtractions) although its effects is marginal, in the few percent range, depending on the choice of scale. Clearly, despite the fact the $gg$ flux is very large it nevertheless has negligible effect on the cross-section simply because the corresponding partonic cross-sections are very small. The large gluon pdf does have a substantial impact on the three-photon cross-section but this happens through the $qg$ reaction. As also emphasized in ref.~\cite{Catani:2018krb} for the case of diphoton production, the $qg$ reaction starts to contribute only at NLO. This leads to a very unique interplay between purely partonic contributions, including their radiative corrections, and partonic fluxes. Specifically, the $q\bar q$ contribution receives a sizable but not huge NLO radiative correction. At NLO this contribution is now dwarfed by the newly generated $qg$ correction which at this point is only LO. At NNLO the $q\bar q$ result gets another significant yet moderate correction, and the $qg$ reaction also receives sizable but reasonable perturbative correction. At NNLO two new channels open - the $gg$ and $qq'$ ones, the latter being much more significant than the former. As also concluded in ref.~\cite{Catani:2018krb} for the case of diphoton production, NNLO is the first order where all large partonic reactions have already been included together with higher-order corrections to the largest ones; therefore one can reasonably expect that from this point on the yet-higher order N$^3$LO corrections to be derived at some point in the future are likely to start showing a more convergent behavior. 

Before closing this section we would like to emphasize that the pattern of scale dependence observed when going from LO through NNLO should not be viewed as anomalous. The fact that scale dependence increases towards NNLO is simply due to the fact that the scale variation at the lower orders is artificially small and that, as explained in this section, at each new order through NNLO new large partonic reactions enter the process thus increasing the overall scale dependance. The arguments given here imply that starting at N$^3$LO the scale variation should start to decrease. This will be very interesting to check in the future. In summary, in our view, the above arguments imply that the scale dependence of the NNLO prediction is likely not artificially small.

\section{Conclusions}\label{sec:conclusions}

In this work we calculate the NNLO QCD correction to three-photon production at the LHC. Our calculation is complete except for the scale-independent part of the two-loop finite remainder which is included in the leading color approximation. We estimate the effect of the missing two-loop contributions. We expect they are phenomenologically insignificant. 

Our calculation is the first NNLO calculation for a $2\to 3$ scattering process. Although the production of colorless final states is not as complicated as a generic $2\to 3$ reaction, we believe that our calculation clearly demonstrates that current computational technology is capable of dealing with the complicated structure of infrared singularities in multi-final state processes. In particular, based on our experience with dijet production (which was computed within the same {\tt STRIPPER} framework as the present calculation) we think that the NNLO computation of three jets at the LHC is feasible. 

An important part of our work is the calculation of the two-loop amplitude $q\bar q\to \3g$ in the leading color approximation. We have expressed the amplitude in a fully analytic form, defined directly in the physical region. To the best of our knowledge this is the first time a 5-point two-loop amplitude has been expressed in a form readily available for phenomenological applications. To this end we had to go beyond simply producing an analytic result that implements the many momentum crossings inherent to processes with many particles in the final state together with the relevant analytic continuations; we have extensively investigated the question of numerical stability and have been able to evaluate the amplitude numerically in about 30k phase-space points with sufficient numerical precision. We would like to stress that this problem is highly non-trivial due to the large size of the amplitude and the large number of independent transcendental functions that appear in it. 

The evaluation of the two-loop amplitude is expensive in terms of CPU time. We have investigated two possibilities to mitigate this problem: one involves specially generated phase-space points that accelerate the convergence of the phase-space integration, while the other involves the construction of a four-dimensional interpolating function which internally utilizes machine-learning techniques. We find that both approaches lead to compatible predictions within the corresponding Monte Carlo errors.

We observe that the structure of higher-order corrections in inclusive three-photon production is very interesting and resembles closely the one known from diphoton production. We find very large higher-order corrections; the NLO prediction for the fiducial cross-section is larger than the LO one by a factor of 2.8 while the NNLO exceeds the NLO one by a factor of 1.6. We have presented detailed analysis of the anatomy of the higher-order corrections in this process and have concluded that the NNLO prediction is likely to be reliable.

Finally, we have compared our predictions with the high-quality LHC data available from the ATLAS Collaboration. We find that the sometimes huge discrepancies between QCD predictions and data noted previously at NLO are absent at NNLO and that the NNLO prediction agrees well with data for all distributions. This result clearly demonstrates how indispensable higher-order corrections are to quantitative phenomenological LHC analyses.

\begin{acknowledgments}
We would like to thank Samuel Abreu, Ben Page and Lance Dixon for illuminating discussions. A.M. thanks the Department of Physics at Princeton University for hospitality during the completion of this work. The work of M.C. was supported by the Deutsche Forschungsgemeinschaft under grant 396021762 - TRR 257. The research of H.C., A.M. and R.P. has received funding from the European Research Council (ERC) under the European Union's Horizon 2020 research and innovation programme (grant agreement No 683211). The work of A.M. was also supported by the UK STFC grants ST/L002760/1 and ST/K004883/1.
\end{acknowledgments}

\end{document}